\documentstyle[12pt]{article}
\begin{document}
\title {\Large {\Large Liouville Field Theory on\\ 
Hyperelliptic surface}}
\author{S.A.Apikyan $^\dagger$ \\Theoretical Physics Departament\\
Yerevan Physics Institute\\ Alikhanyan Br.st.2, Yerevan,375036 Armenia}
\maketitle
\begin{abstract}

Liouville field theory on hyperelliptic surface 
is considered. The partition function of the Liouville
field theory on the hyperelliptic surface 
are expressed as a correlation function of the
Liouville vertex operators on a sphere and the twist fields.

\end{abstract}

\vfill
\hrule
\ \\
\noindent
$\dagger$ e-mail address: apikyan@vxc.yerphi.am\\

\newpage
\indent
{\Large Introduction}\\
\indent
Last years an essential progress has been achieved 
in the investigations of Liouville Field Theory (LFT).
Recently an analytic expression of the three-point
function of he exponential fields in the
Liouville field theory on a sphere is proposed
\cite{DO1},\cite{DO2},\cite{ZZ}. It provides 
possibility for the decomposiion of the multipoint
correlation functions into the multipoint conformal blocks.
Such a success owes much to the fact
that the Liouville correlation functions with exponential
operators inserted in some cases is expressed as a 
multipoint correlation function of vertex operators
in a free field theory \cite{GL}.

The present work is organized as follows. In section I a brief
description is given of a Liouville Field Theory on 
hyperelliptical surfaces. Further by
using Polyakov's proposal \cite{T}, partition function
of the LFT on hyperelliptic surface reduced to LFT 
on sphere and free scalar
field theory with inserted Liouville vertex 
operators and twisted fields.
\\
\\
\\
{\Large1. LFT on hyperelliptic surfaces.}\\
\\
\renewcommand{\theequation}{1.\arabic{equation}}
\setcounter{equation}{0}
\indent
A hyperelliptice surface (HES) $\Gamma_h$ is a compact Riemann 
surface of genus $h\geq1$ determined by an algebraic equation
of the form
\begin{equation}
y^2(z)=P_{2h+2}(z)=\prod_{i=1}^{2h+2}(z-a_{i})
\end{equation}
where $P_{2h+2}(z)$-is a polynomial of degree
$2h+2$. By other words $\Gamma_h$ is a two sheet 
brunched covering of a Riemann sphere.

We start with Liouville field theory (LFT) 
in the conformal gauge on HES $\Gamma_h$.
For the case of the non-zero genus one uses
a global coordinate, provided by the Schottky
uniformization \cite{ZT1},\cite{ZT2}.
Local propeties of the LFT are derived 
from the Lagrangian density
\begin{equation}
{\cal L}=\frac{1}{4\pi}(\partial_a \phi)^2 +
\mu e^{2b\phi} + \frac{Q}{4\pi}\hat{R}\phi 
\end{equation}
where $b$ and $\mu$ are coupling and cosmological 
constants respectively. We have fixed abackground 
metric $\hat{g}$ with curvature $\hat{R}$
normalized by
\begin{equation}
\frac{1}{4\pi}\int \sqrt{\hat{g}}\hat{R}=2(1-h)
\end{equation}
on a genus-h surface.

Under the hyperelliptice map (1.1) Lagrangian density 
${\cal L}(\phi(y))$ and conformal fields $\phi(y),T(\phi)$
on Riemann surface $\Gamma_h$ maps into branches
${\cal L}^{(k)}(\phi^{(k)}(z))$, $\phi^{(k)}(z)$ and
$T^{(k)}(z)$ (i.e. branches of the original fields)
$(k=0,1)$.

Each Liouville field $\phi^{(k)}(z)$ 
(on each sheet separatly) is not exactly 
a scalar field, but transforms like a logarithm 
of the conformal factor of metric under the 
holomorphic change of coordinates
\begin{equation}
\phi^{(k)}(\omega,\bar{\omega})=\phi^{(k)}(z,\bar{z})-
\frac{Q}{2}\log\mid\frac{d\omega}{dz}\mid^2
\end{equation}
where
\begin{equation}
Q=b+\frac{1}{b}
\end{equation}
On each sheet we have holomorphic Liouville
stress-energy tensores
\begin{equation}
T^{(k)}(z)=-(\partial \phi^{(k)})^2 + 
Q \partial^2\phi^{(k)}
\end{equation}
\begin{equation}
\bar{T}^{(k)}(z)=-(\bar{\partial} \phi^{(k)})^2 + 
Q \bar{\partial}^2\phi^{(k)}
\end{equation}
with the Liouville central charge
\begin{equation}
\hat{c}=1+6Q^2
\end{equation}
If $A_a,B_a$ $(a=1,2,...,h)$ denoted the basic 
cycles of the HES, the monodromy operators
$\hat{\pi}_{A_a}$,  $\hat{\pi}_{B_a}$, 
form arepresentation of the $Z_2$ group.
We will work in the diagonal basis of the  
group $Z_2$ i.e. we will choose following 
boundary conditions
\begin{equation}
T=T^{(0)}+T^{(1)},\quad T^-=T^{(0)}-T^{(1)},
\quad \Phi=\phi^{(0)}+\phi^{(1)},
\quad \varphi=\phi^{(0)}-\phi^{(1)}
\end{equation}
\begin{equation}
\hat{\pi}_{A_a} T=T ,\quad
\hat{\pi}_{A_a} T^-=T^- ,\quad
\hat{\pi}_{A_a} \Phi=\Phi ,\quad
\hat{\pi}_{A_a} \varphi=\varphi 
\end{equation}
\begin{equation}
\hat{\pi}_{B_a} T=T ,\quad
\hat{\pi}_{B_a} T^-=-T^- ,\quad
\hat{\pi}_{B_a} \Phi=\Phi ,\quad
\hat{\pi}_{B_a} \varphi=-\varphi
\end{equation}
Under a holomorphic change of coordinates
fields $\Phi$ and $\varphi$ transform as follows
\begin{equation}
\Phi(\omega,\bar{\omega})=\Phi(z,\bar{z})-
Q \log\mid\frac{d\omega}{dz}\mid^2
\end{equation}
\begin{equation}
\varphi(\omega,\bar{\omega})=\varphi(z,\bar{z})
\end{equation}
New expressions for the conserved energy-momentum
tensors are given by
\begin{equation}
T=-\frac{1}{2}(\partial \Phi)^2 + 
Q \partial^2 \Phi -\frac{1}{2}(\partial \varphi)^2
\end{equation}
\begin{equation}
T^-=-\partial \Phi \partial \varphi + 
Q \partial^2 \varphi
\end{equation}
\begin{equation}
\bar{T}=-\frac{1}{2}(\bar{\partial} \Phi)^2 + 
Q \bar{\partial}^2 \Phi -
\frac{1}{2}(\bar{\partial} \varphi)^2
\end{equation}
\begin{equation}
\bar{T}^-=-\bar{\partial} \Phi \bar{\partial} \varphi + 
Q \bar{\partial}^2 \varphi
\end{equation}
and corresponding Liouville central charge is
\begin{equation}
c=2+12Q^2
\end{equation}
The expressions (1.14-1.18)-make obvious 
the splitting of the original theory into the LFT 
on a sphere with the central charge $c_s=1+12Q^2$
and the $Z_2$-orbifold theory for field
$\varphi$ with the central charge $c_{orb}=1$.

We can now define $\Phi$ as Liouville field
on the whole complex plane with the following
asimtotic behavior
\begin{equation}
\Phi(z,\bar{z})=-2Q\log\mid z \mid^2 + reg. 
\quad \mid z \mid \rightarrow \infty
\end{equation}
but $\varphi$-is usual free scalar field
living on the orbifold $S^1/Z_2$ of radius $R$.

According to parity of the boundary conditions
(1.9-1.11) we have to define two kinds of Liouville 
vertex operators. First one (untwisted case)
the spinless primary conformal fields
\begin{equation}
V_{\alpha,\beta}(x)=
e^{2\alpha\Phi (x)}e^{i2\beta \varphi}
\end{equation}
of dimensions
\begin{equation}
\Delta_{\alpha,\beta}=
2\alpha(Q-\alpha)+2\beta^2
\end{equation}
\begin{equation}
\Delta_{\alpha,\beta}^-=
i2\beta(Q-2\alpha)
\end{equation}

As was established in papers \cite{CT},\cite{GN}
the physical LFT space of states consists of a 
continum variety of primery states corresponding 
to operators $e^{2\alpha\Phi}$ with 
\begin{equation}
\alpha=ip+\frac{Q}{2}
\end{equation}
and the conformal descendents of these states.
Therefore the corresponding expression for the
dimensions (1.21) and (1.22) looks as
\begin{equation}
\Delta=\frac{Q^2}{2}+2p^2+2\beta^2 ,\quad
\Delta^-=4p\beta 
\end{equation}
The second one (twisted case) the primary spinless
conformal fields are taken by
\begin{equation}
V_{\gamma}(x)=
e^{2\gamma \Phi(x)}\sigma_{0,\epsilon}(x)
\end{equation}
with dimensions
\begin{equation}
\Delta_{\gamma,\epsilon}=
2\gamma(Q-\gamma)+\frac{1}{16}
\end{equation}
where $\sigma_{0,\epsilon}$ ($\epsilon=0,1$) are well-known twist
(Ramond) conformal fields \cite{Z} of dimension
$\Delta_{0,\epsilon}=1/16$. By definition these
fields $\sigma_{0,\epsilon}$ satisfies the
following nonlocal OPE's:
\begin{equation}
\partial\varphi(z)\sigma_{0,\epsilon}(0)=
\frac{1}{2}z^{-1/2}\sigma_{1,\epsilon}(0) +
\ldots
\end{equation}
\begin{equation}
\partial\varphi(z)\sigma_{1,\epsilon}(0)=
\frac{1}{2}z^{-3/2}\sigma_{0,\epsilon}(0) + 
2z^{-1/2}\partial\sigma_{0,\epsilon}(0) + \ldots
\end{equation}
where $\sigma_{1,\epsilon}$ is second Ramond 
conformal field, dimension for which is defined
from the general formulas of the arbitrary Ramond
fields $\sigma_{k,\epsilon}$ \cite{AZ}
\begin{equation}
\Delta_{k,\epsilon}=
\frac{(2k+1)^2}{16} , \indent k=0,1,2...
\end{equation}

Now we have to answer how can one costract
the partition function of the LFT on $\Gamma_h$
by using vertex operators (1.20),(1.25) 
(with different "charges" $\alpha$,$\beta$,$\gamma$).
According to main proposal of Polyakov \cite{T} the 
"summation" over smooth metries with the 
insertion of vertex operators should be 
equivalent to the "summation" over metries
with singularities of the insertion points,
without insertion of the vertex operators.
Therefore partition function of the LFT on
HES $\Gamma_h$
\begin{equation}
Z_h=\int D\phi e^{-\int\frac{1}{4\pi}
(\partial \phi)^2 + \mu e^{2b\phi}
+ Q \hat{R}^h\phi}
\end{equation}
we can represente as following
\begin{equation}
Z_h=\int D\Phi D\varphi e^{-\int\frac{1}{8\pi}
(\partial \Phi)^2 + \mu e^{2a\Phi}
+ Q \hat{R}^{h=0} \Phi + 
\frac{1}{8\pi}(\partial\varphi)^2}
\prod_{i=1}^{2h+2}e^{2\gamma_i\Phi(a_i,\bar{a}_i)}
\sigma_{0,\epsilon_i}(a_i,\bar{a}_i)
\end{equation}
where
\begin{equation}
a=\frac{Q}{2}\pm \sqrt{\frac{Q^2}{4}-\frac{1}{2}}
\end{equation}
In order to evaluate the last functional
integral, we first integrate over the zero
mode of $\Phi$. Let's define following
decomposition
\begin{equation}
\Phi(z)\equiv\tilde{\Phi}(z)+\Phi_0
\end{equation}
where $\Phi_0$- are the kernel of the Laplacian
and $\tilde{\Phi}(z)$- are functions orthogonal to
the kernel. After integrating over $\Phi_0$ we read
\begin{equation}
\begin{array}{l}
Z_h=\left(-\mu \right)^s \;\frac{\Gamma(-s)}{2a}
\int D\tilde{\Phi} e^{-\int\frac{1}{8\pi}
(\partial \tilde{\Phi})^2 + 
Q \hat{R}^{h=0} \tilde{\Phi}}(\int \sqrt{\hat{g}}
e^{2a\tilde{\Phi}})^s \times\\
\nonumber\\
\qquad \times\prod_{i=1}^{2h+2}e^{2\gamma_i\tilde{\Phi}(a_i,\bar{a}_i)}
\int D\varphi \prod_{i=1}^{2h+2}
\sigma_{0,\epsilon_i}(a_i,\bar{a}_i)
e^{-\int \frac{1}{8\pi}(\partial\varphi)^2}
\end{array}
\end{equation}
where
\begin{equation}
\sum_{i=1}^{2h+2}\gamma_i=Q-sa ,\indent
\sum_{j=1}^{h+1}k_j=0
\end{equation}
The second relation in (1.35) is the condition of
charge's neutrality of intermediate states 
in the pairs OPE's \cite{Z},
\begin{equation}
\sigma_{0,\epsilon_1}(1)\sigma_{0,\epsilon_2}(2)=
\sum z^{-1/8-2(k_{nm}^{\epsilon_1 \epsilon_2})^2}
e^{i2k_{nm}^{\epsilon_1 \epsilon_2}\varphi (2)}
\end{equation}
\begin{equation}
k_{nm}^{\epsilon_1 \epsilon_2}=\frac{n}{R}+
\frac{1}{2}\left[m+\frac{1}{2}(\epsilon_1+\epsilon_2)\right]R
\end{equation}
where $R$-is radius of orbifold.
Unfortunatly, in general,
\begin{equation}
s=\frac{Q}{a}-\frac{\sum\gamma_i}{a}
\end{equation}
will not be a positive
integer, therefore remaining functional integral in
(1.34) is not a free-field correlator. But
every time when (1.35) is satisfied for intiger 
$s=n=0,1,2,...$ partition function $Z_h$ exhibits a pole in the
$\sum \gamma_i$ with the residue being specified 
by the corresponding perturbative integral
\begin{equation}
\mathrel{\mathop {res}\limits_{\sum\gamma_i=Q-na,\,\sum k_j=0}} 
Z_h(a_1,...,a_{2h+2}) =
\left.Z_h^{(n)}(a_1,...,a_{2h+2})\right|_
{\sum\gamma_i=Q-na,\,\sum k_j=0}
\end{equation}
where $Z_h^{(n)}$-is free field correlator
\begin{equation}
\begin{array}{l}
Z_h^{(n)}=\frac{\left(-\mu \right)^n}{n!}\;
\int D \Phi e^{-\int\frac{1}{8\pi}
(\partial \Phi)^2 + 
Q \hat{R}^{h=0} \Phi}(\prod_{j=1}^{n}\int \sqrt{\hat{g}}
e^{2a\Phi(x_j)}d^2x_j) \times\\
\nonumber\\
\qquad \times\prod_{i=1}^{2h+2}e^{2\gamma_i \Phi(a_i,\bar{a}_i)}
\int D\varphi \prod_{i=1}^{2h+2}
\sigma_{0,\epsilon_i}(a_i,\bar{a}_i)
e^{-\int \frac{1}{8\pi}(\partial\varphi)^2}
\end{array}
\end{equation}
which is $n$-th term in the perturbative series of $Z_h$
\begin{equation}
Z_h(a_1,...,a_{2h+2})=\sum_{n=0}^{\infty}\;Z_h^{(n)}(a_1,...,a_{2h+2})
\end{equation}
expanded by the cosmological constant $\mu$.
So LFT's partition function on HES reduced to 
Liouville correlation function (on the sphere)
with inserted Liouville vertex operators
(with charges $\gamma_i$) and to
correlation function of twisted fields
$\sigma_{0,\epsilon}$ \cite{Z}. The residue 
of the LFT's partition function on HES at pole 
$\sum\gamma_i=Q-na$ is a correlation 
function of free field theory on HES.
\\
\\
\\
{\bf Acknowledgments} I would like to thank
G.Mussardo, A.Sedrakian and I.Vaysburd
for
useful discussions. This work was completed
during my visit in the ICTP, Trieste. I thank
this institute for the hospitality.
\\
\\
\\
\\

\end{document}